# Evidence for a temperature dependent anisotropy of the superconducting state parameters in underdoped SmBa$_2$Cu$_3$O$_x$

A. Kortyka,[1,2] R. Puzniak,[1] A. Wisniewski,[1] M. Zehetmayer,[2] H. W. Weber,[2] C. Y. Tang,[3] X. Yao,[3] and K. Conder[4]

[1] Institute of Physics, Polish Academy of Sciences, Aleja Lotników 32/46, PL-02-668 Warsaw, Poland

[2] Atominstitut, Vienna University of Technology, 1020 Vienna, Austria

[3] Department of Physics, Shanghai Jiao Tong University, 800 Dungchuan Road, Shanghai 200240, P. R. China

[4] Laboratory for Developments and Methods, Paul Scherrer Institute, CH-5232 Villigen PSI, Switzerland

The temperature dependence of the anisotropy of the superconducting state parameters, $\gamma$, was studied by torque magnetometry for the high temperature superconductor SmBa$_2$Cu$_3$O$_x$ in magnetic fields of up to 9 T. The measurements were performed on four underdoped single crystals with oxygen contents corresponding to $T_c$'s varying from 42.8 to 63.6 K. The anisotropy was found to be strongly temperature dependent, while only a weak dependence on the magnetic field was observed. No evidence for a field dependent superfluid density was found. Possible origins of the temperature dependence of the anisotropy are discussed.

**I. Introduction**

A lot of attention has been paid in the last decade to the temperature dependence of the effective mass anisotropy in layered superconductors, which was initiated by the discovery of the temperature dependence of the upper critical field anisotropy in $MgB_2$ (Ref. 1) and followed by its interpretation as being the consequence of the two-band nature of classical *s*-wave superconductivity.[2-4] Recently, a temperature dependence of both the penetration depth[5] and the upper critical field[6] anisotropy was reported for the pnictide superconductors, thus reviving the discussion on multiband scenarios and the mechanisms leading to superconductivity with high $T_c$.[7-9] However, until now, the issue of a temperature dependence of the effective mass anisotropy in the best known class of the layered high-$T_c$ superconductors (HTSC), i.e. the cuprate superconductors, has never been seriously treated. This may be caused by the very weak experimental indications of multigap superconductivity in this class of materials. However, the multigap mechanism is not the only possibility leading to a temperature dependence of the anisotropy parameter. The motivation for the present work was to probe the temperature dependence of the anisotropy parameter in the cuprates and to try to answer the question if such dependence is a common intrinsic property of all layered superconductors with high transition temperature.

The simplest description of anisotropy in layered superconductors relies on the classical anisotropic Ginzburg-Landau theory (AGLT), where the anisotropy is introduced via the anisotropy parameter of the effective mass of the superconducting carriers $\gamma = \sqrt{m_c^*/m_{ab}^*} = \lambda_c/\lambda_{ab} = H_{c2}^{\|ab}/H_{c2}^{\|c} = \xi_{ab}/\xi_c$.[10] Here, $m_{ab}^*$ and $m_c^*$ are the effective charge carrier masses related to supercurrents flowing in the *ab*-planes and along the *c*-axis, respectively; $\lambda_{ab}$ and $\lambda_c$ are the corresponding penetration depths, $H_{c2}^{\|ab}$ and $H_{c2}^{\|c}$ are the upper critical fields, and $\xi_{ab}$ and $\xi_c$ are the corresponding coherence lengths. The above theory assumes a single band anisotropic system and temperature and field independent effective masses. Furthermore, AGLT does neither account for the occurrence of an in-plane anisotropy in the tetragonal basal plane nor for the positive curvature of $H_{c2}(T)$. The first breakdown of a description within the AGLT approach was reported for niobium.[11] The non-spherical Fermi surface was proposed to be responsible for the observed temperature dependence of the upper critical field anisotropy.[12] It was shown later that the anisotropy of the electron-phonon interaction and of the Fermi velocity can indeed explain the superconducting properties of Nb.[13,14] A large (for a cubic material) anisotropy of $H_{c2}$ was reported for $V_3Si$.[15] Moreover, anisotropy effects were observed in the basal plane of $Cs_{0.1}WO_{2.9}F_{0.1}$ with sixfold symmetry, which fails to agree with AGLT that predicts an isotropic effective mass tensor.[16] In both materials, the anisotropy effects were suggested to be due to the shape and anisotropy of the Fermi surface.[15,16] Temperature dependent anisotropy parameters and a pronounced upward curvature of $H_{c2}(T)$ were reported for $NbSe_2$ and $LuNi_2B_2C$ and originally attributed to nonlocal effects.[17,18] A nonlocal relation[19,20] between the current and the vector potential provides a formal bridge between the Fermi system of electrons in a given crystal and the interacting vortices in the superconducting condensate.[21] Such effects may be relevant for high purity samples. Initially, the borocarbides had been considered as simple $s$-wave superconductors[22] and the analysis of $H_{c2}(T)$ using an anisotropic single gap model in $LuNi_2B_2C$ resulted in anisotropic electron-phonon coupling and an anisotropy of the Fermi velocity.[23] Recently, strong support for multiband superconductivity in $LuNi_2B_2C$ was provided[24] which should be followed by a re-examination of the previous analysis. An unusually strong temperature dependence of the anisotropy parameter was found for $MgB_2$ (Refs 1 and 2). An anisotropic single gap model, which results in an incorrect gap anisotropy of the order of 10, could not explain the temperature dependence of the anisotropy parameter, see Ref. 25 and references therein. It was suggested that the existence of different gaps was responsible for the temperature dependence of the anisotropy.[2,25-28] Recently, strong evidence for a temperature dependent anisotropy and indications for multiband superconductivity were reported for the iron-based superconducting pnictides.[5,6,29-31] The complicated electronic structure involves at least four energy bands near the Fermi surface.[32] Despite great progress, more research is needed to clarify the origin of the temperature dependence of the anisotropy parameter in this class of compounds. Taking the above into account, the general question can be raised whether or not the temperature dependence of the anisotropy parameter is a common feature of a larger group of superconductors, particularly also the HTSC.

Clear evidence that the anisotropy parameter is temperature dependent or independent is still missing for cuprates. A broad temperature range was investigated for $Y_2Ba_4Cu_8O_{16}$ and a temperature independent anisotropy suggested.[33] However, the shaking technique was not employed in these studies performed on samples with a pronounced fishtail effect. Vortex shaking is based on the application of an additional oscillating magnetic field perpendicular to the main field, which pushes the vortices from weak pinning centers and extends the reversible region in the ($H,T$) phase diagram.[34] Optimally doped crystals of $YBa_2Cu_3O_x$, Sr-doped $YBa_2Cu_3O_x$, and mercury-based HTSC were studied only in quite a narrow temperature range.[35-37] Occasionally, the shaking technique was used[36] to reduce vortex pinning and no evidence for a temperature dependent anisotropy parameter was found. This failure, nevertheless, may be due to several factors: (i) pronounced pinning, sometimes not fully eliminated by vortex shaking, making the derived anisotropy uncertain, (ii) too narrow temperature range, (iii) anisotropy in optimally doped crystals that may in fact be temperature independent, which does not have to be the case for underdoped systems.

Measurements of the anisotropy of HTSC in a wide temperature range are experimentally challenging. One of the most accurate methods to measure the anisotropy parameter is torque magnetometry. This technique has been used successfully, e.g., to investigate transitions of the vortex structure for different orientations of the applied field or to explore new metamagnetic states.[38,39] Despite its broad applicability, torque magnetometry is not adequate for determining the mechanisms leading to a temperature dependent anisotropy, i.e. whether or not cuprates are multiband superconductors, but can provide accurate data on the anisotropy parameter. Spectroscopic investigations, probing the superconducting gaps directly, are needed to clarify the multiband scenario.[40,41] To make the torque technique reliable for anisotropy investigations, measurements should be performed in magnetic fields, where the hysteresis is small or negligible, i.e., at $H$ of the order of the irreversibility field in the $ab$-plane, $H_{irr}^{\|ab}$, and in fields below $H_{c2}^{\|c}$. For highly anisotropic superconductors with irreversibility lines at high magnetic fields, it is often difficult to fulfill these conditions in a wide temperature range. These problems are mitigated in strongly underdoped $REBa_2Cu_3O_x$ superconductors (RE123, RE – a rare earth or Y), i.e., with highly reduced oxygen content, $x$, where the upper critical field and the irreversibility field are strongly reduced. An example of such superconductors is $SmBa_2Cu_3O_x$ (Sm123).

The relevance of the 2D Lawrence-Doniach (LD) model[42] should be examined when studying highly anisotropic HTSC with strongly reduced $H_{c2}^{\|c}$. The LD model views layered superconductors as a stacked array of planes coupled by Josephson tunneling, whereas AGLT describes them as a continuous 3D medium characterized by the anisotropy parameter $\gamma$ introduced above. To fully justify the applicability of the AGLT approximation, $\xi_c(T)$ has to be larger than the interplanar distance $d$, which is of the order of 0.8 nm for RE123.[43] Otherwise, the magnetic torque for fields almost parallel to the $ab$-plane should be described by the 2D approach. In reality, deviations from the continuous medium London approach are due to the scaling function $\varepsilon(\theta)$, see below, and become only important for angles $\theta < \theta_0 = \tan^{-1}(1/\gamma)$,[44] i.e., for angles of the order of 1° away from the $ab$-plane geometry in the case of HTSC. Moreover, it was shown that the 3D London model successfully applies to the highly anisotropic Bi-based and Hg-based superconductors[37,45] with higher anisotropy than that in the compounds being the subject of the current work. Therefore, the AGLT approach was applied for the analysis of the data obtained in the present studies.

Here, we report on torque magnetometry measurements of the anisotropy of the superconducting state parameters in underdoped $SmBa_2Cu_3O_x$ single crystals with various $T_c$'s. We determined the anisotropy from the reversible torque and found that $\gamma$ is strongly temperature dependent. Section II presents details of the sample preparation and of the measuring technique. In Sec. III, a short description of the data evaluation is presented and support for a temperature dependence of the anisotropy parameter is derived. Additionally, some of the possible scenarios explaining the anisotropy behavior are discussed. In Sec. IV conclusions are drawn.

## II. Experimental details

Single crystals of Sm123 were grown by top seeded solution growth.[46] Several crystals with similar plate-like geometry and masses of about 2 mg were selected for our studies. They were annealed[47] in flowing oxygen-helium gas at various temperatures between 490 and 505 ºC and oxygen partial pressures between 0.009 and 0.4 bar in order to obtain various oxygenation levels. Their $T_c$ was determined by ac susceptibility measurements performed with an amplitude of 0.1 mT and a frequency of 10 kHz in a 9 T Physical Property Measurement System (Quantum Design, PPMS). The values of $T_c$ were found to be 42.8 K, 51.5 K, 56.5 K, and 63.6 K for the crystals denoted S1, S2, S3, and S4, respectively. The difference in $T_c$ confirms the variation of the intentionally introduced oxygenation level, see inset of Fig. 1a and Table I. The XRD analysis (D-5000 Siemens diffractometer using Cu K$\alpha$ radiation) confirmed the crystals' high quality and was used to determine their lattice constants, see Table I.

The magnetic torque measurements were performed in the PPMS equipped with a torque option, in the temperature range from about 10 K below $T_c$ up to $T_c$ and in magnetic fields of up to 9 T. The temperature range was limited by torque hysteresis, which becomes pronounced at low temperatures, see below. A small sinusoidal normal state background of anisotropic paramagnetic origin was subtracted from all data.

## III. Results and discussion

The magnetic torque, $\tau$, was recorded for increasing and decreasing angles $\theta$ between the $c$-axis of the crystal and the applied magnetic field over an angular range of 180º, in steps of 0.5º. Some examples of the torque, measured for all the crystals, are presented in Fig. 1a.

The investigated crystals show very weak pinning, which is due to the very long annealing time between 100 and 620 hours and makes the torque nearly reversible in the full angular range. The free energy of an anisotropic superconductor in the reversible regime of the mixed state for fields $H_{c1} \ll H \ll H_{c2}$ was calculated by Kogan *et al.*[49-51] within the 3D anisotropic London model approach. The corresponding angular dependence of the superconducting torque in the reversible region is given by the first term on the right-hand side of the following expression:

$$\tau(\theta) = -\frac{V\Phi_0 H}{16\pi\lambda_{ab}^2}\left(1-\frac{1}{\gamma^2}\right)\frac{\sin(2\theta)}{\varepsilon(\theta)}\ln\left(\frac{\eta H_{c2}^{\|c}}{\varepsilon(\theta)H}\right) + A\sin(2\theta). \quad (1)$$

Here, $V$ is the volume of the crystals, $\Phi_0 = 2.07\times10^{-15}$ Tm$^2$ is the flux quantum, $\eta$ is a numerical parameter of the order of unity depending on the structure of the flux-line lattice, and $\varepsilon(\theta) = \left[\cos^2(\theta) + \gamma^{-2}\sin^2(\theta)\right]^{1/2}$. The second term on the right-hand side describes the contribution of an anisotropic paramagnetic or diamagnetic susceptibility and can be treated as a background contribution to the torque in the superconducting state,[52,53] with $A$ describing the amplitude of the background torque. By measuring the angular dependence of the torque in the mixed state of a superconductor with anisotropic paramagnetic or diamagnetic background, four parameters can be extracted from the data: the in-plane magnetic penetration depth, the *c*-axis upper critical field, the effective mass anisotropy, and the background torque amplitude. A sinusoidal background does not affect $\gamma$ significantly, see Fig. 1b, in contrast to the effect on $H_{c2}^{\|c}$, which may easily differ by 50%. Therefore, it is reasonable to fix $H_{c2}^{\|c}$ using $H_{c2}^{\|c}(T)$ values obtained from another method (see below) thus reducing the number of fit parameters. The first term on the right-hand side of Eq. (1) describes the reversible torque calculated from $\tau_{rev}(\theta) = \left(\tau(\theta^+) + \tau(\theta^-)\right)/2$ obtained by clockwise, $\tau(\theta^+)$, and counterclockwise, $\tau(\theta^-)$, rotating the sample in the magnetic field, see the insets of Fig. 2a and 2b. The values of $H_{c2}^{\|c}$ were fixed in the fitting procedure using $H_{c2}^{\|c}(T)$ from magnetization measurements in a 7 T SQUID (Quantum Design, MPMS), see Table I. An example of such a magnetization curve is shown in the inset of Fig. 1b. The values of $H_{c2}^{\|c}(0)$, as shown in Table I, were obtained from the dependence introduced in Ref. 49, which was later presented in a more useful form (assuming the clean limit) as $H_{c2}(0) = 0.7255\times T_c\times dH_{c2}/dT$ (Ref. 10). It was already shown[5] and confirmed by our experiment, that small errors in the $H_{c2}$ values do not much affect the anisotropy parameter extracted from the fit of Kogan's equation for the angular torque dependence. Therefore, we set, as is commonly done in such an analysis, $\eta = 1$. The three other parameters of Eq. (1) were extracted simultaneously in the torque fitting procedure. An example of raw torque data before and after subtracting the background and the background contribution is shown in Fig. 1b for the sample with $T_c = 56.5$ K.. It would be of interest to compare the anisotropy values extracted from the torque for $T \rightarrow T_c$ with those for the anisotropy of $H_{c2}$ near $T_c$. However, for strongly underdoped samples, the anisotropy is very sensitive to the oxygen content. Therefore, it seems to be impossible to compare results obtained on samples annealed under different conditions, i.e. those presented here with those published elsewhere concerning the anisotropy of $H_{c2}$. The large anisotropy makes the crystal's orientation for $H$ along the $ab$-plane critical in deriving the exact $\gamma$ value, hence the $\gamma$ values obtained from magnetization or transport measurements may be underestimated. No attempts to obtain the upper critical field anisotropy directly from magnetization measurements were made because of the large upper critical fields, the large anisotropy, and the paramagnetic background contribution (which could be avoided e.g. in the case of $MgB_2$). The values of the upper critical field can be derived by applying torque magnetometry for determining $H_{c2}(\theta)$ (Ref. 1), but for crystals with a paramagnetic background contribution the $H_{c2}(\theta)$ values would be highly inaccurate, as mentioned above.

Sometimes the measurements had to be performed in applied fields close to $H_{c2}^{\|c}$, i.e. for $H > 0.6 H_{c2}^{\|c}(T)$, to overcome significant irreversibility at angles near 90°. In those cases the fit of Eq. (1) to the torque was performed in a reduced, e.g. 60° < $\theta$ < 120°, angular range to fulfill the condition $H \ll H_{c2}$. No significant curvature of $H_{c2}^{\|c}(T)$ was observed in the vicinity of $T_c$, which is common for cuprates in fields applied along the uniaxial axis.

The temperature and the field range for the angular torque investigation was chosen in such a way as to be, on one hand, as broad as possible keeping $H(T) < H_{c2}(T)$ and, on the other hand, to minimize the torque hysteresis at angles close to 90°, i.e., for the $H\|ab$-plane. We wish to point out that in all torque measurements a nearly reversible signal was obtained for clockwise and counterclockwise rotating the crystal in the full angular range. Even for the crystal with the strongest pinning, i.e. with the highest $T_c = 63.6$ K, a nearly reversible torque was recorded, see the inset of Fig. 2a. This makes errors due to averaging of the torque negligible and the derived anisotropy parameter highly reliable. Excellent fits of Eq. (1) to the torque data were obtained, see the example in the inset of Fig. 2b.

The measurements performed in a constant magnetic field show an increase of the anisotropy parameter with decreasing temperature. This behavior was found for all Sm123 crystals studied, see Fig. 2a. For the crystal with the lowest $T_c = 42.8$ K, the anisotropy, recorded in a magnetic field of 2 T, increases from 35.9 ± 0.8 to 59.2 ± 1.2 while lowering the temperature from 40 to 36 K. For Sm123 with the highest $T_c$ = 63.6 K, the observed increase in the same magnetic field is from 13.6 ± 0.2 to 21.8 ± 0.1 upon decreasing the temperature from 62 to 59 K. The increase of the anisotropy parameter amounts to over 50 % in both cases. Due to the very different upper critical field values among the crystals, a comparison of the temperature dependence of the anisotropy parameters can be presented at constant $H/H_{c2}^{\|c}$ values, see Fig. 2b. The anisotropy parameter decreases with increasing temperature and appears to be in first approximation a linear function of temperature with a slope that depends on the $T_c$ of the crystal. This slope changes from -8.6/K to -0.5/K for the crystals with $T_c$ increasing from 42.8 to 63.6 K. Additionally, in order to verify that the temperature dependence of the anisotropy parameter is not influenced by thermal fluctuations, data points in Fig. 2c are presented at constant $T/T_{c2}^{\|c}$ values. Here, $T_{c2}^{\|c}$ is the temperature of the superconducting-to-normal state transition in a magnetic field applied along the crystallographic *c*-axis. Again, a clear dependence of the anisotropy parameter on the temperature is visible. An example of the reversible torque at two different temperatures/fields with the fits of Eq. (1) is presented in the inset of Fig. 2c. Such a comparison provides evidence for an increase of the anisotropy parameter with decreasing temperature. Taking the above into account, we conclude that the temperature dependence of the anisotropy parameter is an intrinsic property of Sm123. The anisotropy increases with decreasing temperature for underdoped Sm123 and appears to depend more strongly on temperature for the crystals with lower $T_c$, i.e. for more strongly underdoped crystals.

In contrast to the clear temperature dependence, the dependence of the anisotropy parameter on the magnetic field is rather weak. The anisotropy decreases somewhat for higher magnetic fields at the same temperature, see Fig. 3. Nevertheless, since this effect is not so pronounced, we cannot rule out that it results from systematic errors coming e.g. from a systematic change of the hysteresis width with magnetic field. Strong pinning close to the *ab*-plane affects the positions of the peaks in the reversible (averaged) torque and hence makes the derived anisotropy parameter uncertain. The weak decrease of the anisotropy parameter with increasing magnetic field may as well be due to the reduced angular range of the fit of Eq. (1) to the data, when the field approaches $H_{c2}^{\|c}$, see the inset of Fig. 3. The anisotropy parameter remains rather constant at small fields and decreases only when the field becomes comparable with $H_{c2}^{\|c}$, at around 0.6 $H_{c2}^{\|c}$, i.e. at fields where the reduced angular range of the fit of Eq. (1) to the data, was performed. Therefore, it is possible that the weak field

dependence of the anisotropy parameter may not be an intrinsic property of Sm123, but rather a systematic error when approaching $H_{c2}^{||c}$.

The superfluid density, $\rho_s$, can be probed directly by measuring the magnetic penetration depth via $\lambda_{ab}^{-2} \propto \rho_s$. The field dependence of the superfluid density in Sm123 was investigated by extracting the values of $\lambda_{ab}$ from the torque measurements. Representative torque data collected on two Sm123 crystals in various magnetic fields are presented in Fig. 4 with the angular dependence approximated by Eq. (1). The in-plane penetration depth does not show a significant variation with magnetic field, see the representative data in the insets of Fig. 4. The analysis, like all measurements, was limited to those temperatures and fields, where the torque does not show a pronounced irreversibility.

The lack of a field dependence of the superfluid density reported here remains in contrast to the behavior observed in MgB$_2$ (Refs 54-56), in the pnictides,[31] and in La$_{1.83}$Sr$_{0.17}$CuO$_4$ (Ref. 57). A two-gap model was applied for MgB$_2$ and the pnictides, and a ($d + s$)-wave gap symmetry, i.e. a two-band scenario, was suggested for the cuprates. In the two-gap model, the field dependence of $1/\lambda_{ab}^2$ is due to a sum of two superconducting band contributions with two different superfluid densities to the total superfluid density. The strong suppression of superconductivity with increasing magnetic field in one band with only the large gap surviving in strong magnetic field, leads to the field dependent superfluid density. It was pointed out, that superconducting gaps, characterized by different symmetries, show different field dependences, i.e. the suppression of the superfluid density with magnetic field in a $d$-wave gap is proportional to $H^{1/2}$ (Ref. 58) while it has a $1/H^{1/2}$ dependence[59] in an $s$-wave gap. Since a multiband scenario, responsible for the field dependence of $\rho_s$, is not the only possibility leading to a temperature dependent anisotropy, all possible situations will be analyzed in the following.

In principle, the origin of the observed temperature dependence of the anisotropy parameter in Sm123 may be related to one of at least five situations: (i) multiband superconductivity,[60-69] (ii) Fermi surface anisotropy,[70-72] (iii) unconventional pairing and anisotropy of the superconducting energy gap,[73-80] (iv) strong coupling,[81-83] (v) real limitations of AGLT in the case of highly underdoped superconductors due to their strictly layered structure. Nonlocality, which becomes observable when the mean free path becomes larger than the superconducting coherence length, may be a necessary ingredient for the situations (i)-(iii).

Multiband superconductivity was proposed as an extension of the conventional BCS theory[60] and the phenomenon of two-gap superconductivity was observed in several systems.[61,62] The temperature dependence of the anisotropy parameter may be explained by the existence of two topologically very different Fermi surfaces, as in $MgB_2$, where two anisotropy parameters, namely of the penetration depth and of the upper critical field, were distinguished.[27,63] The cylindrical Fermi surface of $MgB_2$, that is dominant at low temperature/high magnetic field, gives a large anisotropy, while the $\pi$ band, due to its much larger Fermi velocity along the $c$-axis, plays a more important role at temperatures close to $T_c$, i.e. at low magnetic field, and, therefore, reduces the upper critical field anisotropy.[25] Some implications for multiband superconductivity in the cuprates appeared[8,9,64], but were not confirmed to date. To our knowledge, there are so far no direct spectroscopic indications for a second superconducting gap in the energy spectrum.[65] Strong electron-electron correlations make those materials quite difficult to treat with first principle calculations.[66] The Fermi surface in the cuprates[67] is not yet fully understood and only recently an unambiguous observation of quantum oscillations in the Hall resistance of underdoped Y123 proved the existence of a well developed Fermi surface.[68] It was found that $\lambda_{ab}^{-2}$ decreases with increasing magnetic field for $MgB_2$ and the pnictides[31,55], both materials for which a temperature dependent anisotropy parameter was reported. An increase of the anisotropy with increasing field was found in $MgB_2$ (Ref. 54), whereas the anisotropy parameter was found to be field independent, at least up to 1.4 T, in the pnictides.[69] In both superconductors, mixing of two superconducting bands explains very well the field dependence of the anisotropy parameter and of the superfluid density.[31,54] No evidence for a field dependent penetration depth was found in Sm123, however, a scenario of multiple band or multiple gap superconductivity in the cuprates cannot be excluded completely based on our findings. For Sm123, the observed behavior of $1/\lambda_{ab}^2$ may be due to the range of applied fields, which was dictated by the torque hysteresis and the $H_{c2}^{\|c}$ values. Firstly, too small fields might not be large enough to suppress superconductivity in any of the bands. On the other hand, the smallest field used in this work was 1 T (in order to make the torque signal sufficiently strong), which might be already too large, as the suppression of one of the superconducting gaps was visible for fields below approximately 0.3 T in the case of $La_{1.83}Sr_{0.17}CuO_4$ (Ref. 57). More research on the superfluid density in underdoped cuprates is needed, in order to investigate possible similarities between the behavior of the cuprates and $MgB_2$ or the pnictides and to shed more light on the hypothesis of multiband superconductivity in the cuprates.

Detailed Fermi surface structures are essential to describe the upper critical field in type-II superconductors, as first noticed by Hohenberg and Werthamer[12] and shown later explicitly.[70,71] It was demonstrated[71] that, as the Fermi surface approaches the Brillouin zone boundary, i.e. the Fermi surface changes from almost spherical to highly distorted due to the crystal symmetry, the dimensionless upper critical field parameter $h(t) = H_{c2}(t)/(-dH_{c2}(t)/d(t))$, where $t = T/T_c$, is much enhanced in comparison with the value for the isotropic model. The calculations presented by Kita and Arai[71] clearly indicate that the Fermi surface anisotropy can be the main source of the upward curvature in $H_{c2}$ near $T_c$ and, therefore, may explain the temperature dependence of the anisotropy parameter. When the Fermi surface anisotropy is fully taken into account in energy band calculations, the numerical results excellently reproduce the experimental $H_{c2}(T)$ values of conventional superconductors.[72] Due to the lack of knowledge of the Fermi surface structure in the underdoped cuprates, it is impossible to eliminate this scenario as the origin of the temperature dependence of the anisotropy parameter in Sm123.

Isotropic s-wave superconductivity cannot lead to a temperature dependence of the upper critical field anisotropy.[73] Conversely, d-wave pairing even in a superconductor with an isotropic Fermi surface can result in an upward curvature of $H_{c2}(T)$ near $T_c$. An additional temperature dependence of the anisotropy parameter is obtained when assuming an anisotropic effective mass.[73] The gap symmetry in the cuprates appears not to be a pure d-wave[74-77] and indications of a mixed, i.e. (d + s), symmetry in the cuprates have been concluded by many groups.[76,78,79] It has been shown so far, that the anisotropy of the superconducting order parameter can lead to changes in anisotropy with temperature of the order of 20 %.[80] At present, the scenario, in which the temperature dependence of the anisotropy parameter observed in Sm123 results from the anisotropy of the superconducting gap, must probably be included, although more investigations are needed to arrive at a final conclusion.

The main effect of having included strong-coupling corrections to $H_{c2}(T)$ is to alter the electron effective mass, i.e. the Fermi velocity, from the band mass value.[81] This produces an increase in $H_{c2}$ compared to using the bare band Fermi velocity, but the shift being relatively temperature-independent cancels out of $h(T)$.[82] A strong polaron coupling approach, proposed as a further extension of the BCS phonon-mediated superconductivity for the cuprates,[82,83] could possibly be included in this scenario. However, when detailed calculations are made,[83] strong coupling gives only minor modifications to $H_{c2}(T)$ and cannot explain the observed strong temperature dependence of the anisotropy parameter.

For $MgB_2$ and the pnictides, for both of which a temperature dependent anisotropy parameter was reported, two different anisotropy parameters, i.e. the anisotropy of the upper critical field $\gamma_{Hc2}$ and of the penetration depth $\gamma_\lambda$, were invoked following Kogan's approach.[5,54,84] For Sm123, with no significant field dependence of the anisotropy, this would imply the same temperature dependence for both $\gamma_{Hc2}$ and $\gamma_\lambda$. No peculiarities in the angular dependence of the torque would suggest that $\gamma_{Hc2}$ and $\gamma_\lambda$ would not differ much. However, a direct determination of $H_{c2}$ and of its anisotropy from the angular dependence of the torque was not possible. Furthermore, additional independent anisotropy measurements that would indicate different anisotropies, i.e. of the penetration depth and the coherence length, are lacking. Therefore, it would be highly speculative at present to suggest that the temperature dependence of $\gamma$ may originate from two anisotropy parameters.

Further studies, especially of the superconducting gap symmetry and the Fermi surface of underdoped cuprates, are highly desirable to arrive at final conclusions. Possibly there might be a mixing of different anisotropy effects involved that would lead to the observed strong temperature dependence of the anisotropy in the underdoped high temperature superconductor Sm123. In particular, situations (ii) and (iii) may occur together, and (i) may be just an extreme case of the combination of (ii) and (iii). The effect may also have quite a different origin. It may very well be that the temperature dependence of the anisotropy of the superconducting state parameters is much more common than had been expected so far. A temperature independent effective mass anisotropy is one of the basic assumptions of AGLT, leading, for a single gap superconductor, to a temperature independent anisotropy of the penetration depth and of the coherence length. However, this may not be true for highly underdoped superconductors. For a strongly layered superconductor with Josephson coupled planes, a reduction of the interlayer coupling with temperature would imply an increase in the anisotropy parameter. According to the LD model, $\xi_c = \xi_{ab}(m_{ab}/m_c)^{1/2} \propto \xi_{ab} t_\perp$, where $t_\perp$ is the interlayer coupling constant, which was found to be temperature dependent in a 2D system.[85] Therefore, it may be necessary to reconsider the temperature dependence of the interlayer coupling in highly underdoped cuprates and to formulate a new theory describing the anisotropy in strongly layered HTSC.

## IV. Conclusions

A careful study of the anisotropy of the superconducting state parameters in underdoped Sm123 single crystals with $T_c$ varying from 42.8 to 63.6 K was performed. The effective mass anisotropy parameter was found to be temperature dependent for all of the investigated Sm123 single crystals. This effect is indeed intrinsic. In contrast to the strong temperature dependence, only a very weak dependence of the anisotropy parameter on magnetic field was observed, but it cannot be excluded that this is the result of systematic errors caused by performing measurements at fields close to $H_{c2}^{\|c}$. No dependence of the superfluid density on the magnetic field was found. Since no detailed information on the Fermi surface of underdoped cuprates is available, the origin of the observed temperature dependent anisotropy parameter remains unclear. Our work shows that underdoped cuprates, besides the multiband superconductors, belong to the same group of superconductors, where the temperature dependence of the anisotropy parameter is an intrinsic property.

**Acknowledgements**

The critical reading of the manuscript and the comments made by M. Angst are gratefully acknowledged. We wish to thank V. Domukhovski for the x-ray measurements. A. K. thanks the European NESPA project for financial support. This work was partially supported by the Polish Ministry of Science and Higher Education under the research projects No. N N202 4132 33 and N N202 2412 37. X. Y. thanks the Shanghai Committee of Science and Technology Grants.

Table I. Abbreviations for the investigated Sm123 crystals, their transition temperatures, lattice constants, upper critical field slopes, and zero temperature upper critical fields.

| Sample | $T_c$ (K) | $a, b, c$ (nm) | $\mu_0 dH_{c2}^{\|c}/dT|_{T_c}$ (T/K)[a] | $\mu_0 H_{c2}^{\|c}(0)$ (T)[b] |
|---|---|---|---|---|
| S1 | 42.8 | 0.38619(4), 0.39092(4), 1.17738(2) | -0.24 | 7.5 |
| S2 | 51.5 | 0.38573(1), 0.39131(2), 1.17585(1) | -0.45 | 16.8 |
| S3 | 56.5 | 0.38533(3), 0.39131(2), 1.17491(1) | -0.6 | 24.6 |
| S4 | 63.6 | 0.38516(2), 0.39131(2), 1.17438(1) | -1.05 | 48.5 |

[a] from magnetization measurements
[b] assuming clean limit and WHH dependence[48]

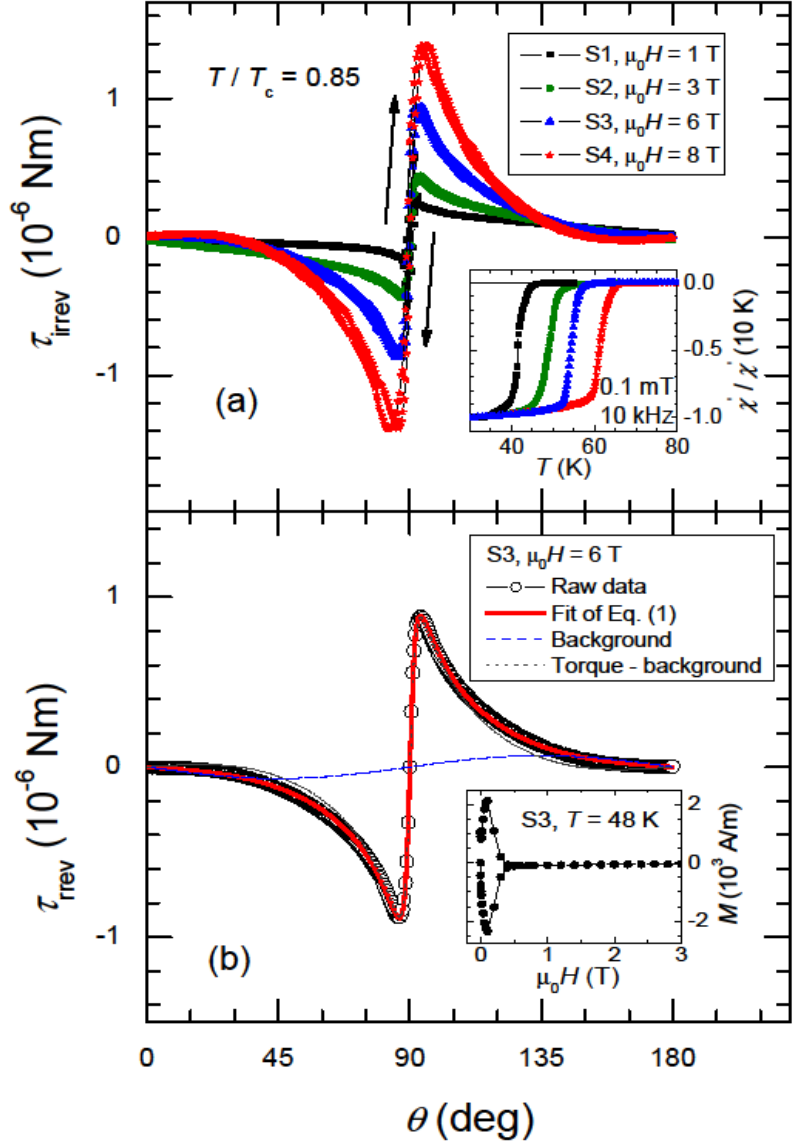

Fig. 1. (Color online) a) Clockwise and anticlockwise angular dependence of the torque for all crystals recorded at the same reduced temperature $T/T_c = 0.85$ in different magnetic fields. b) Averaged reversible torque for the crystal with $T_c = 56.5$ K (S3) compared with the fit of Eq. (1). The background and the superconducting torque contributions are presented as well. Insets: a) Temperature dependence of the real part of the ac susceptibility at an amplitude of 0.1 mT and a frequency of 10 kHz. b) Example of the field dependence of the magnetization for the crystal with $T_c = 56.5$ K (S3).

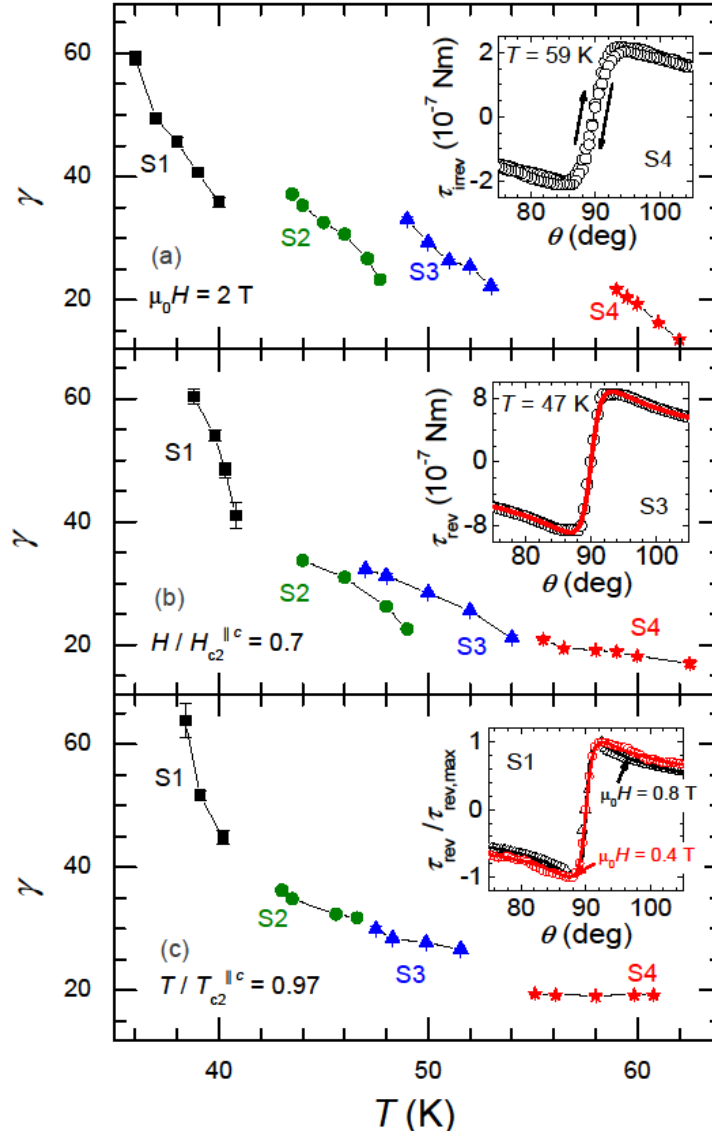

Fig. 2. (Color online) Temperature dependence of the anisotropy parameter a) at $\mu_0 H = 2$ T, b) at a reduced field $H/H_{c2}^{\parallel c}$ of 0.7, c) at a reduced temperature $T/T_{c2}^{\parallel c}$ of 0.97. Insets: a) Example of the clockwise and anticlockwise angular dependence of the torque at $\mu_0 H = 2$ T. b) Example of the reversible torque at $H/H_{c2}^{\parallel c} = 0.7$ fitted with Eq. (1). c) Example of the reduced reversible torque at $T/T_{c2}^{\parallel c} = 0.97$ fitted with Eq. (1). Here, $T_{c2}^{\parallel c}$ is the temperature of the superconducting-to-normal state transition in a magnetic field applied along the crystallographic $c$-axis.

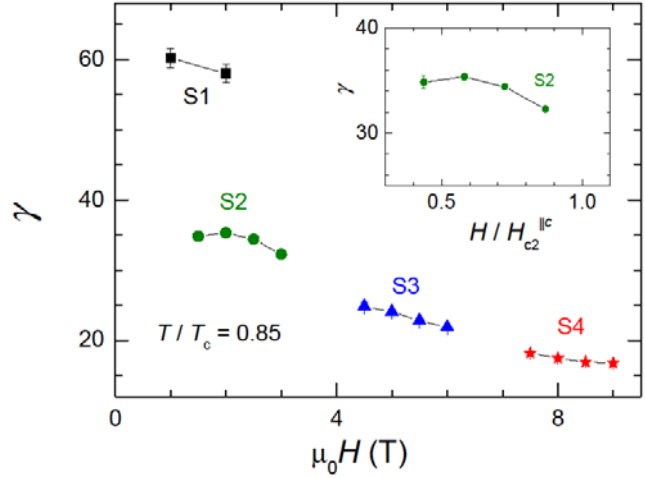

Fig. 3. (Color online) Field dependence of the anisotropy parameter at fixed reduced temperature. Inset: Reduced field dependence of the anisotropy parameter for the crystal with $T_c = 51.5$ K (S2).

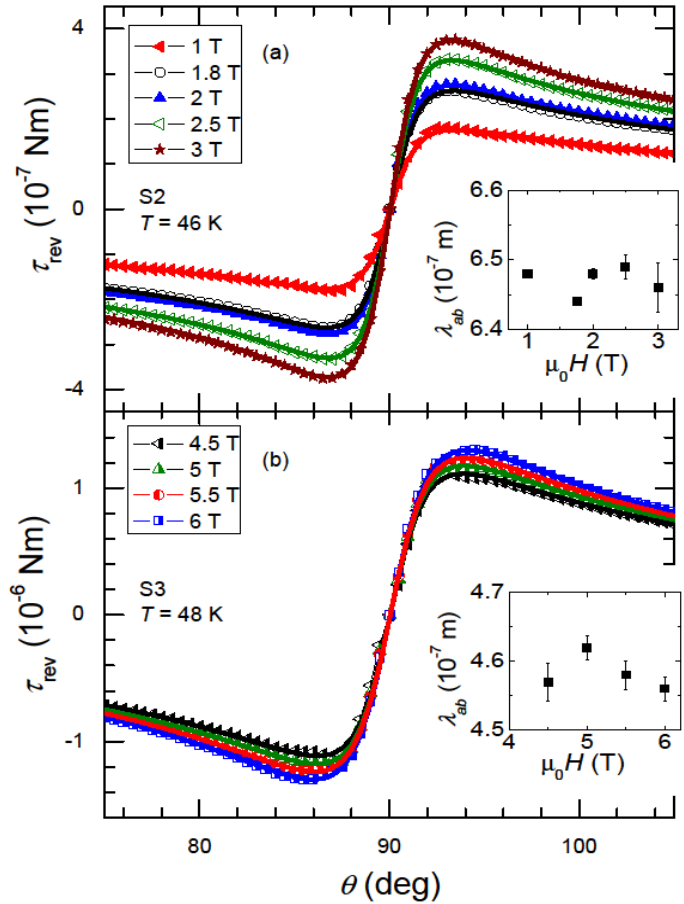

Fig. 4. (Color online) Reversible angular torque for the crystals with $T_c = 51.5$ K (S2) and with $T_c = 56.5$ K (S3) fitted by Eq. (1), panel a) and b), respectively. Insets: a) Example of the field dependence of the penetration depth for S2; b) same for S3.